\documentclass[]{spie}  

\usepackage{amsmath,amsfonts,amssymb}
\usepackage{wrapfig}
\usepackage{graphicx}
\usepackage[colorlinks=true, allcolors=blue]{hyperref}
\usepackage{subfig}
\usepackage{tikz}
\usetikzlibrary{calc,trees,positioning,arrows,chains,shapes.geometric,%
    decorations.pathreplacing,decorations.pathmorphing,shapes,%
    matrix,shapes.symbols}
\tikzstyle{block1} = [rectangle, draw, fill=blue!20, text width=10em, text centered, minimum height=2em]
\tikzstyle{block2} = [rectangle, draw, fill=red!20, text width=10em, text centered, minimum height=2em]
\tikzstyle{line} = [draw, -latex']
\tikzstyle{cloud} = [draw, ellipse,fill=red!20, node distance=5cm,  minimum height=2em]
\tikzstyle{block3} = [draw, ellipse,fill=green!20,  minimum height=2em]
\tikzstyle{block4} = [draw, rectangle,fill=yellow!20,  minimum height=2em]

\title{On the Exclusive Jet Measurements at the LHC}

\author[a]{Paula Erland}
\affil[a]{Institute of Nuclear Physics PAS, Radzikowskiego 152, 31-342 Cracow, Poland}

\authorinfo{Further author information: \\Paula Erland: E-mail: paula.erland@ifj.edu.pl}

\begin{document} 
\maketitle

\begin{abstract}
Properties of the exclusive jet production process are briefly described. 
A need for a dedicated trigger algorithms for studies of such 
process in the LHC experiments is explained. Influence of the machine luminosity leveling -- a change of the betatron function and 
crossing angle -- on the scattered proton position in the AFP is discussed. Finally, impact of pile-up and optics on the expected algorithm 
efficiency is estimated.
\end{abstract}


\section{Introduction}
\label{sec:intro}  

\begin{figure}[h!]
\centering
\includegraphics[width=0.25\textwidth]{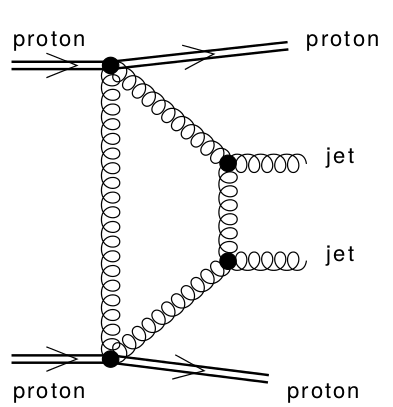}
\caption{Leading Feynman diagram of the exclusive jet production: two gluon jets, two scattered protons and a presence of a ``screening'' gluon to make exchange colourless.}
\label{fig:EXC}
\end{figure}

In this paper some aspects of the exclusive jet measurement at the LHC are discussed. Exclusive jet production (see Feynman diagram in Figure \ref{fig:EXC}) is a very rare and interesting 
process, which is still quite poorly known\cite{exc_jets_general}. It can be described by:
\begin{itemize}
\item two jets: usually produced centrally (at small rapidity),
\item two protons: scattered at small angles (few hundreds $\mu$rad from the beam axis),
\item rapidity gap between the central system and the scattered protons.
\end{itemize}
A term ``exclusivity'' means that all final state particles can be measured. This implies a strong relation of the kinematics of the central state and the scattered protons. High luminosity at the LHC gives a hope to significantly measure properties of the exclusive jet production in ATLAS \cite{ATLAS} and CMS/TOTEM \cite{CMS, TOTEM} experiments. In this paper a stress will be put on the ATLAS detector but the conclusions should remain valid also for CMS.

Jets can be detected by the ATLAS central detector but the forward scattered protons escape its acceptance region through the beam pipe opening. This underlines a need for a dedicated set of detectors installed far away from the interaction point. For the ATLAS interaction point (IP1) these are the ATLAS Forward Proton (AFP) \cite{AFP} detectors.

AFP system consists of two Roman pot stations (Near and Far) on each side of ATLAS. They are located symmetrically on both sides of the IP1 at the distance of about 210 m from it. The detectors are horizontally inserted into the LHC beam-pipe up to few millimetres from the beam. Detectors located on the ATLAS C side are positioned in the vicinity of  beam 1 whereas the ones on the A side close to beam~2. Each station contains four Silicon Tracker planes (SiT), which provide a precise measurement of the scattered proton trajectory position (spatial resolution of a single plane was measured to be about  6 $\mu$m  and about 30 $\mu$m in $x$ and $y$
respectively \cite{AFP_testbeam}). Far stations host also the Time of Flight (ToF) detectors. The main goal of ToF modules is to reduce the backgrounds, via the measurement of the arrival time of a proton combined with the interaction vertex reconstruction. The timing resolution is expected to be of about 30 ps \cite{AFP_testbeam}.

For obvious reasons the LHC magnetic lattice influences the scattered proton trajectory and one has to take into account the following elements:
\begin{itemize}
  \item two dipole magnets (D1-D2) for beam separation (bending),
  \item five quadrupole magnets (Q1-Q5) for beam focusing,
  \item two collimators (TCL4, TCL5) for magnet protection.
\end{itemize}
Settings of these elements, called the accelerator optics, come from  the requirements of the experiments in terms of luminosity and  the LHC machine protection and may differ between the runs. The properties of various LHC optics are discussed in Ref. \cite{opt}.gnuplot plot sqrt from column in bash script
 
In order to fulfil the exclusivity criteria all stable particles should be measured. For the exclusive jet production, both scattered protons need to be within the AFP acceptance \cite{acceptance}. It means that the energy of the jet system is expected to be between about 300 and 1300 GeV. As was shown in \textit{e.g.} Ref. \cite{ATLAS_jets}, the cross section for such jets is expected to be about few $\mu$b. This means taking the data in the high pile-up conditions, \textit{i.e.} in a situation when multiple $pp$ interactions occur during one bunch crossing.

\begin{figure}[h!]\centering
\subfloat[\label{1bg}]{\includegraphics[width=0.2\textwidth]{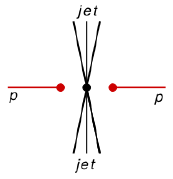}}
\subfloat[\label{2bg}]{\includegraphics[width=0.2\textwidth]{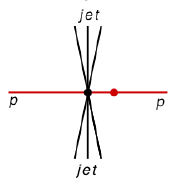}}
\subfloat[\label{3bg}]{\includegraphics[width=0.2\textwidth]{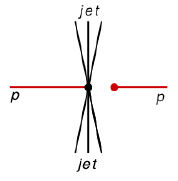}}
\subfloat[\label{4bg}]{\includegraphics[width=0.2\textwidth]{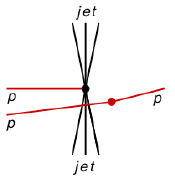}}
\caption{Backgrounds to the exclusive jet production: (a) non-diffractive jets + 2 protons from 2 different pile-up vertices, (b) non-diffractive jets + 2 protons from the same pile-up vertex, (c) single diffractive jets + proton from pile-up, (d) single diffractive jets + proton from pile-up.}
\label{BG}
\end{figure}

As was discussed in \cite{ATLAS_EXC_Jets}, the main gnuplot plot sqrt from column in bash scriptbackgrounds for the exclusive jet production are due to a non-diffractive and single diffractive processes overlaid with scattered protons originating from the minimum-bias pile-up events. The most common backgrounds are shown in Figure \ref{BG}.

The cross-section for the non-diffractive jet production is the highest one among the processes with a hard scale -- the jets are copiously produced in ATLAS. Due to the limited trigger and data acquisition bandwidths not all of these events can be saved for analysis. In ATLAS, events are selected at two trigger steps called level-1 (L1) and High Level Trigger (HLT)\cite{ATLAS}. Due to the storage limitation certain thresholds were introduced at the L1 and HLT stages. Events not fulfilling these criteria are prescaled, \textit{i.e.} only a fraction of them is saved. For jets, the selection is usually based on the minimum transverse momentum, (p$_T$), of a  jet. For data taken in years 2015-2018 (the so-called Run2 period) these thresholds were set to 100 GeV at L1 and about 440 GeV at HLT \cite{ATLAS_jets}. For the future data taking (starting from Run3), these limits may be even higher. Since the cross-section for exclusive jets is very small and steeply falling with increasing jet p$_T$, it is necessary to accept all events being within the detector acceptance. Since, as was discussed above, this  results in jets with p$_T$ staring from 150 GeV, a dedicated trigger algorithm has to be used. In the this paper the focus will be on the HLT items. 

Considered algorithm consists of two parts: 
\begin{enumerate}
  \item comparison of the ``proton'' position computed from the di-jet system kinematics to a real proton position as registered in AFP,
  \item comparison of the jet vertex and the one reconstructed using the AFP Time-of-Flight system.
\end{enumerate}

The idea of the first part is to match the position of ``proton'' which is calculated using the proton properties inferred from the jet kinematics to that of the proton actually registered in the AFP station. The jet observables ($p_T$, energy, rapidity and azimuthal angle) are measured by ATLAS. From all jets in the event two with the highest $p_T$ are selected. At this step the vertex is not requited as it is expected that high $p_T$ jets should not originate from two independent vertices.

Using the jet observables the di-jet system mass ($M_{jj}$) and rapidity ($y_{jj}$) are computed. Next, a relative energy loss ($\xi_{A,C}$) of a proton scattered into the ATLAS A/C hemisphere is calculated as:
$$\xi_{A,C}=\exp(\pm y_{jj}) \frac{M_{jj}}{2 E_{beam}}.$$
The estimated properties of a hypothetical scattered proton are used to simulate its transport through the LHC magnetic lattice and to calculate its trajectory position in the AFP station. Next, this hypothetical position ($p_{jet}$) and the registered one ($p_{proton}$) are compared. If the predicted position is close enough to the measured one, the event is accepted by the trigger. In presented work the default selection criteria used in the current ATLAS trigger algorithm were chosen. It was required that the difference between the two positions in $x$($y$) direction are smaller than 2(2.5) mm and the radius $R = \sqrt{(p^{x}_{proton} - p^{x}_{jet})^2 + (p^{y}_{proton} - p^{y}_{jet})^2} < 2$ mm. Therefore, the effective cut is the one on the radius. These criteria could be optimized to reject as much of the background as possible while keeping  majority of the signal.

The second part of the exclusive jet trigger uses the comparison of the interaction vertex position calculated from  the AFP measured times of flight of the scattered protons and that of the jet vertex reconstructed in ATLAS. A ``proton vertex'' can be computed as: $Z_p=(t_A - t_C)\frac{c}{2}$ and then compared to the jet system related one: $Z_p - z_{jet}$.

The first part of this trigger algorithm strongly depends on the changes of the  LHC optics or the jets kinematics. In this paper, a discussion about the influence of the LHC optics and presence of pile-up is presented.

\section{Impact of the LHC Optics}

In order to collect a large amount of data required  by physics analysis, the LHC tries to maximize the luminosity. There are few methods to achieve this. One uses the relation of the betatron function at IP, ($\beta^*$), and the instantaneous luminosity. The luminosity is inverse proportional to the $\beta^*$ value. Another method hinge on the crossing angle levelling ($\theta_C$). It is an angle at which the beams are crossing at the interaction point\footnote{In this paper the ``crossing angle'' is denoted as and angle between the nominal $z$ beam axis and the $z$ direction of the beam. In literature such angle is sometimes called ``half crossing angle''.}. The range of possible changes is limited by the long range beam-beam separation (required to ensure good beam lifetimes) and by the mechanical aperture of the beam pipe. In order to keep the luminosity high, the LHC decided to adjust both $\beta^*$ and $\theta_C$ during the data taking.

In order to study the levelling influence on the reconstruction of the proton position two $\beta^*$ values, typically used during Run2, were considered: $\beta^*=30$ cm and $\beta^*=40$ cm. It was checked that the resulting proton positions for a given crossing angle are exactly the same. Therefore, all further analysis are done only for $\beta^*=30$~cm.

\begin{figure}[ht]\centering
\includegraphics[width=0.49\textwidth]{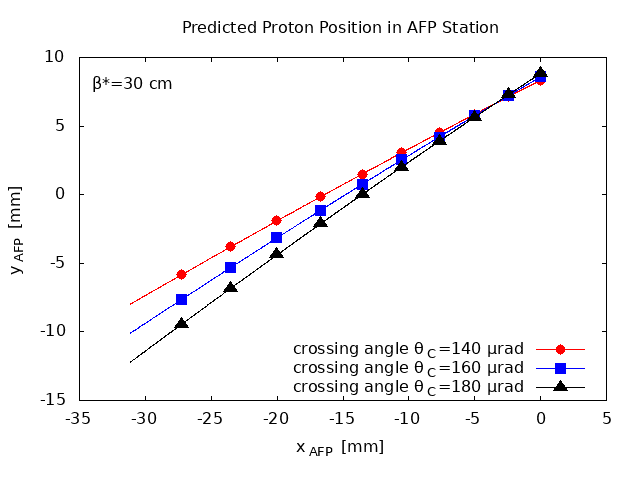}\hfill
\includegraphics[width=0.49\textwidth]{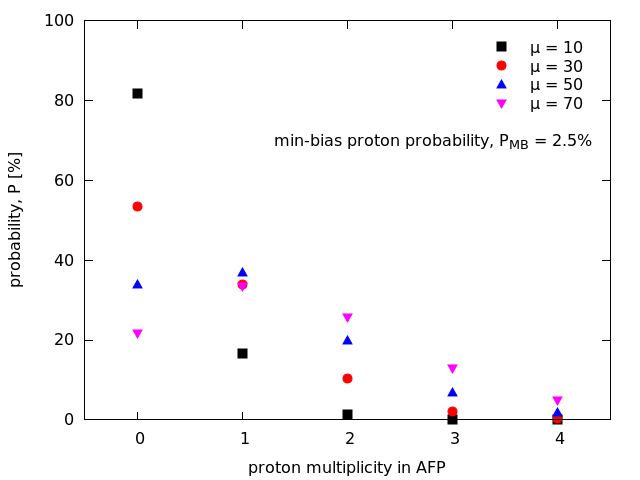}
\caption{\textbf{Left:} proton position in the AFP detector for $\beta^*$=30 cm and various values of crossing angle: 140, 160 and 180 $\mu$rad. \textbf{Right:} proton multiplicity in the AFP detector for various pile-up conditions. The probability of registering a minimum-bias 
proton in the AFP is assumed to be of 2.5\%.}
\label{proton_position}
\end{figure}

In contrast to the $\beta^*$ levelling, a significant differences are observed while applying a change of $\theta_C$. This is clearly visible in Figure \ref{proton_position} (left) where the proton positions in the AFP for the three different values of the crossing angle (140, 160 and 180 $\mu rad$) are shown. Except from the focal point located around $x_{AFP} = -4$ mm, which is expected to be independent of the machine optics\cite{dynam_alg_metod}, the change of $\theta_C$ moves the protons along $y$-axis. At -20~mm from the beam centre $(0,0)$, which is typically an edge of the AFP detector acceptance at high mass, the difference is about few millimetres. This shows the needed to make trigger algorithm optics-dependent.

\section{Impact of Pile-up}
In this section the impact of the pile-up environment on the presence of the scattered protons in AFP and its influence on the jet kinematics is briefly discussed.

The presence of additional protons increases the chance of an event to be accepted by the trigger. The multiplicity of protons visible in AFP for a given pile-up is shown in Figure \ref{proton_position} (right). The plot was prepared under the assumption that a chance of the minimum-bias proton registration in the AFP is 2.5\%. A more detailed discussion about the probability of having the scattered protons in the AFP detector can be found in \textit{e.g.} Ref. \cite{Rafal_doktorat}. As can be seen from the figure, for pile-up of 10 the chance to have more than one proton is negligible. However, at $\mu = 30$ it increases to about 15\% and at $\mu = 50$ it is of about 30\%. This can decrease the background rejection power due to combinatorics as the chance that one of two (three, ...) registered protons will match the one from jet kinematics increases.

Another consequence of the non-zero pile-up environment is its impact on the jet reconstruction. In order to study such effect Monte Carlo samples containing signal (exclusive jets; generated by FPMC\cite{FPMC}), background (non-diffractive jets; generated by Pythia8\cite{Pythia8}) and pile-up (minimum bias events; generated by Pythia 8) were prepared. The jets were reconstructed using the  anti-$kT$ algorithm implemented in the FastJet tool\cite{FastJet}. Two values of the jet radius were used: R = 0.4 and 1.0. The choice was to match the most popular settings used by the ATLAS Collaboration for jet analyses.

\begin{figure}[htbp]
\centering
\includegraphics[width=0.49\textwidth]{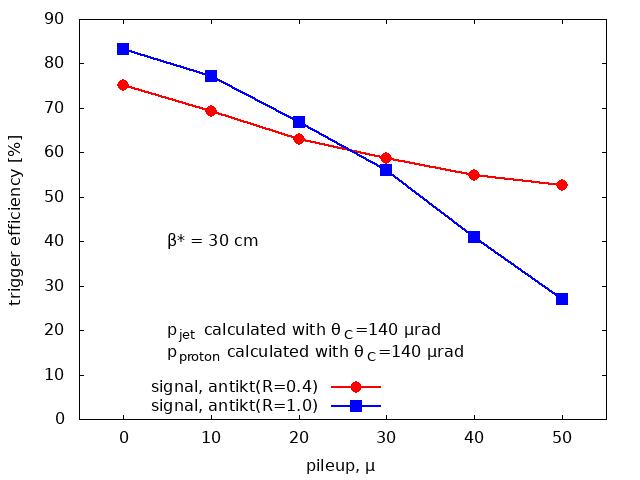}\hfill
\includegraphics[width=0.49\textwidth]{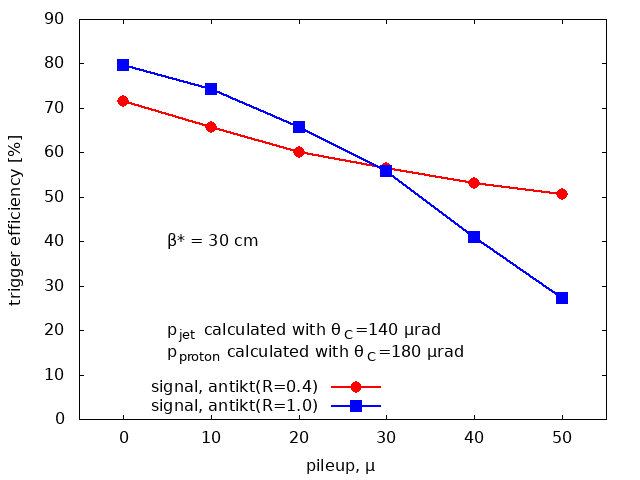}\\
\includegraphics[width=0.49\textwidth]{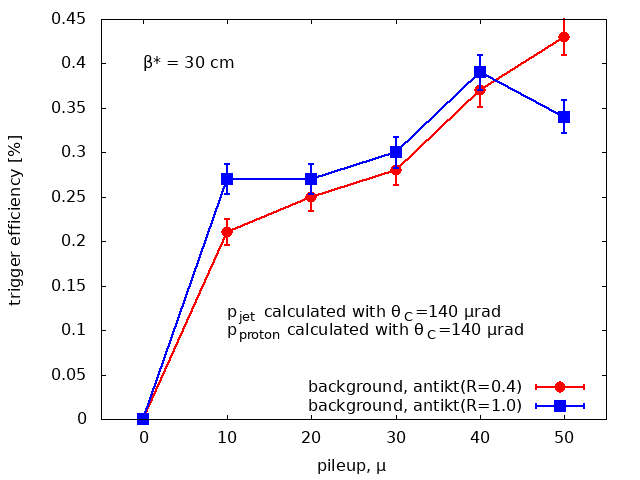}\hfill
\includegraphics[width=0.49\textwidth]{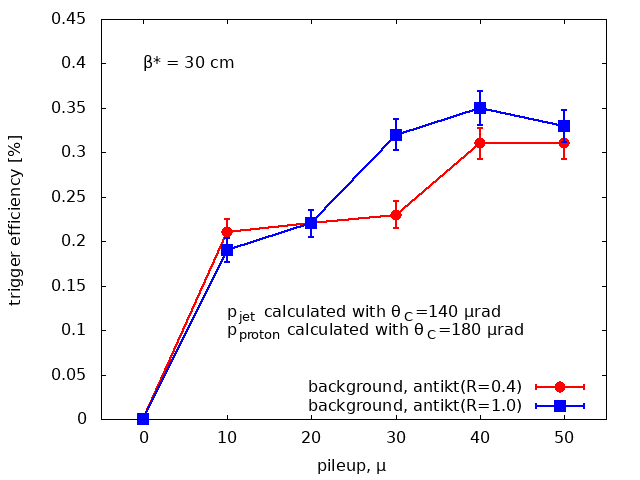}
\caption{Efficiency of HLT exclusive jet trigger algorithm for signal (exclusive jets, \textbf{top}) and background (non-diffractive jets, \textbf{bottom}) as a function of pile-up for two settings of anti-$kT$ algorithm. \textbf{Left} plots are for exact optics. On the \textbf{right} plot ``real'' protons were transported with $\theta_C$ of 180 $\mu$rad whereas ``jet protons'' with $\theta_C = 140$ $\mu$rad which reflects effect of luminosity leveling.}
\label{fig:gggg}
\end{figure}

The trigger efficiency, \textit{i.e.} the number of events triggered by the used algorithm to the number of all events having the double AFP tag, as a function of the pile-up value is shown in Figure \ref{fig:gggg}. Proton was assumed to be tagged if its x position in AFP was smaller than -2.5 mm, which reflects Run 2 data-taking conditions. Top plots are for the signal and bottom ones for the non-diffractive jet background. The value of pile-up was fixed, \textit{i.e.} no Poisson smearing was applied. The selection criteria result the pure selection efficiency (\textit{i.e.} at $\mu = 0$) of about 75\% for anti-$kT$ with $R = 0.4$ and 80\% for $R = 1.0$. The difference is due to the presence of a third (fourth, ...) jet in the event which influences the properties of kinematics of  a ``proton'' reconstructed from jets. The pile-up interactions cause more particles to be present in a registered event. In consequence, the jets have usually higher $p_T$ and slightly different rapidity w.r.t. the $\mu = 0$ situation. In case of signal this significantly worsens the efficiency, especially in the case of anti-$kT$ with $R = 1.0$. In case of background, slightly more events are selected at higher pile-up values. The major effect is the presence of multiple protons (see Fig. \ref{proton_position} (right)). Nevertheless, the rejection power stays at the level of few permille, typically  below 0.5\%. The fluctuations are of a statistical nature (the background sample contained 100 000 events).

It is worth mentioning that ATLAS and CMS Collaborations provided special algorithms \cite{ATLAS_PU_suppression} for the suppression of the pile-up effects on the jet reconstruction. It is clear that such a correction is a must also for the exclusive jet trigger, but such studies are beyond the scope of this paper.

\section{Summary}
The data delivered by the LHC provided a chance to measure properties of the exclusive jet production process. With the current experimental set-up, this can be done by ATLAS and CMS/TOTEM Collaborations, since both jets and scattered protons should be measured. In case of ATLAS, a dedicated trigger algorithms at the  L1 and HLT levels have to be used.

One of important HLT algorithms uses a correlation between the jet and the scattered proton kinematics. Due to the fact that protons are measured about 200 m from the interaction point, they have to travel through the LHC magnets which have a profound impact on their trajectories. This is taken into account in the proton tracking tools.

In this paper, it is shown how a new feature of the LHC machine -- the luminosity levelling -- impacts the trigger efficiency. A change of $\beta^*$ seems to have no effect, contrary to the change of the crossing angle of the beams. In such a case, when $\theta_C$ is changed from 140 $\mu$rad to 180 $\mu$rad, the efficiency drops by few percent. At zero pile-up the efficiency is  about 75\% for anti-$kT$ with $R = 0.4$ and 80\% if $R = 1.0$ is used. Simultaneously, non-diffractive background is strongly suppressed -- only less then 0.5\% of events is accepted.

Presence of high pile-up greatly modifies the signal efficiency since it changes the jet kinematics. Hence, the use of the pile-up suppression algorithms in the jet reconstruction is a must. Its effect on the trigger algorithm should be studied in detail by both Collaborations.

\section*{Acknowledgements}
This work has been partially supported by the Polish Ministry of Science and Higher Education under the ``Diamentowy Grant'' programme (0138/DIA/2017/46). The author would like to thank dr Maciej Trzebi\'nski and prof. Janusz Chwastowski for suggestions, discussions and comments.

\bibliography{report} 
\bibliographystyle{spiebib} 

\end{document}